\documentclass[10pt]{article}
\usepackage{latexsym,graphicx,amssymb,color,amsmath,bbm}
\usepackage{epsfig,lscape}

\def\Z{\mathbb{Z}}
\def\N{\mathbb{N}}
\def\half{\frac{1}{2}} 
\def\t{\tau} 

\begin{document}
\vspace*{-2cm}
\begin{flushright}
\end{flushright}

\vspace{0.3cm}

\begin{center}
{\Large {\bf A toolkit for the construction of icosahedral  \\ \vspace{0.3cm}
particles with local symmetry axes }}\\ 
\vspace{1cm} {\large \bf R.\ Twarock\,\footnote{\noindent E-mail: 
{\tt rt507@york.ac.uk}}}\\
\vspace{0.3cm} {\em Department of Mathematics and Department of Biology \\ University of York\\
York YO10 5DD, U.K.}\\ 
\end{center}

\begin{abstract}
A formalism is developed which allows to determine the locations of all local symmetry axes of three-dimensional particles with overall icosahedral symmetry. It relies on the fact that the root system of the non-crystallographic Coxeter group $H_3$ encodes the locations of the planes of reflection that generate the discrete rotational symmetries of the particles.  Via an appropriate extension of the root system, new planes of reflection are introduced which determine local  axes of rotational symmetry. An easy-to-implement formalism is derived that allows to compute the surface structure of any three-dimensional icosahedral particle with local symmetries. It can be used also for particles with overall octahedral and tetrahedral symmetry  in conjunction with the root systems of the corresponding reflection groups. 

Applications to viruses are discussed explicitly. It is shown that the concept of quasi-equivalence in Caspar-Klug Theory corresponds to the special case of local six-fold symmetry axes contained in the theory developed here, and the corresponding geometries can hence be obtained with this formalism based on the root system of $H_3$. 

Moreover, as a by-product, the theory answers the long-standing open question why only certain types of capsomeres, i.e. clusters of protein subunits, are observed in the surface structures of viruses. Since the types of the capsomeres are determined by the orders of the local symmetry axes on which they are located, the possible types of capsomeres are restricted by the spectrum of local symmetry axes allowed by the theory. Based on this we determine the spectrum of all capsomere types that may occur in viral capsids and give explicit examples for the lower-order cases. 
\end{abstract}

\section{Introduction}

Icosahedrally symmetric particles with local symmetries play a crucial role in virology, because viruses encapsulate their genome in protein containers, called viral capsids, that are organised in protein clusters centred on the symmetry axes of the capsids. These axes of symmetry may be global, i.e. corresponding to rotations that leave the whole structure invariant and are described by the overall group structure, or local, in which case only the local environments around the symmetry axes. An example of a local symmetry is the discrete rotation about the centre of the capsomere, that leaves only the capsomere itself, but not the whole capsid, invariant. 
The classification of all possible local symmetries of icosahedral surface structures hence determines the spectrum of possible viral capsids with icosahedral symmetry. 

Viruses with icosahedrally symmetric capsids occur predominantly in nature \cite{Casjens:1985}. By making use of symmetry, a virus can efficiently encode the structure of its capsid: it only needs to encode the locations of the protein subunits modulo symmetry, that is only the locations of a subset of the proteins, and the locations of all other proteins follow by symmetry. Since the icosahedral group is the finite symmetry group of largest order in three dimensions (order 60, as opposed to order 48 for the octahedral group, and order 24 for the tetrahedral group) it is hence not surprising that icosahedral symmetry plays a distinguished role.\footnote{It allows to compress the data necessary for encoding of the locations of the protein subunits in the capsid by a factor of 1/60.} 

If only the symmetry axes of the icosahedral group (12 5-fold axes, 20 3-fold axes and 30 2-fold axes) could be used to encode the locations of protein subunits, only small viral capsids would occur. Viral capsids therefore also follow local symmetry axes. This has already been observed by Caspar and Klug \cite{Caspar:1962}, and they classify the structure of particles that follow a particular type of local symmetry, which runs under the terminology {\it quasi-equivalence} in their work. However, experimental evidence has shown that viruses follow also local symmetries different from these \cite{Casjens:1985}. Inspired by this, a mathematical formalism is developed here which can be used to determine the surface structures of all icosahedral particles with local symmetries.  

The approach adopted here is based on the fact that the rotational symmetries of the icosahedral group are generated by the reflections that form the non-crystallographic Coxeter group $H_3$. The locations of the corresponding planes of reflection are encoded by a set of vectors, called root vectors, that are the (normalized) vectors perpendicular to the reflection planes. In order to obtain further planes of reflection that correspond to local rotational symmetries and are compatible with the overall group structure (global symmetries), the set of root vectors has to be extended in an appropriate way. In order to achieve this, affine extensions of non-crystallographic Coxeter groups are used \cite{Twarock:2002AffH} as discussed in section \ref{two}. They lead to an augmentation of the set of roots by a further root, and $\Z$-linear combinations of this enlarged set of roots define nested point sets that are subsets of generalised lattices. The points in these sets correspond to the normal vectors of additional reflection planes and hence encode the locations of additional (local) symmetry axes that are compatible with the overall group structure. 
The structures of icosahedral particles with local symmetries can thus be reconstructed based on an affine extension of the non-crystallographic Coxeter group $H_3$. In section \ref{two} a simple tool-kit is developed based on this formalism which can be used to compute the structure of these particles. It is easy to implement and can be used also by people who are not familiar with group theory. Applications are considered in section \ref{three}. 

It is important to note that the formalism presented here can be used to construct {\it all} icosahedral particles with local symmetry axes. The reason for this lies in its connection with the projection formalism known from the study of quasicrystals \cite{Senechal:1996,Shechtman:1984} and Penrose tilings \cite{Penrose:1974}. In particular, it is known that point sets with local symmetries can be obtained via a projection from a higher dimensional regular lattice such as, for example, the root lattices corresponding to indecomposable finite root systems \cite{King:2004}. All local symmetries of particles with overall icosahedral symmetry can hence be obtained via projection from the root lattice of $D_6$  \cite{Kramer:1987}. 
Since the root system of $H_3$ can be obtained from the root system of $D_6$ by projection \cite{MoodyPatera}, and the root lattice of $D_6$ is given by $\Z$-linear combinations of the root vectors, which after projection lead to $\Z$-linear combination of the root vectors of $H_3$, one can equally work with $\Z$-linear combinations of the root vectors of $H_3$ themselves provided that one has a guide as to which linear combinations are important. This information is provided by the affine extension of the group, and leads to the tool-kit for the construction of particles with local symmetries presented in subsection \ref{twotwo}.  

Therefore, even though one could in principle construct the surface structures of the particles via projections from the root lattice of $D_6$, a different approach is adopted here, which is less cumbersome because  remnants of the  higher dimensional  space are only implicitly contained in the theory. All practical computations reduce to very simple operations in three dimensions that can easily be implemented on a computer. Applications are discussed in section \ref{three}. 

\section{The group theoretical set-up}\label{two}

Due to the importance of icosahedral symmetry in virology, we present the theory with focus on this case. However, the theory can be used also for the derivation of the surface structures of particles with octahedral and tetrahedral symmetry, if the root systems of the corresponding reflection groups are used instead of the root system of $H_3$. 

\subsection{Background}\label{twoone}

The Coxeter group $H_3$ is the only non-crystallographic Coxeter group in three dimensions. It is a subgroup of $H_4$, the only non-crystallographic Coxeter group in four dimensions, and contains $H_2$ as a subgroup \cite{Humphreys:1992}. The root systems of $H_4$ and $H_2$ can be obtained via projection from the root systems of the crystallographic groups $A_4$ and $E_8$, respectively. As in the case of $H_3$ and $D_6$, the higher dimensional crystallographic root systems in these pairings are distinguished by the fact that all root vectors are of the same length. 

The elements of $H_3$ are reflections $r_j$ at planes perpendicular to the root vectors (or roots in short) $\alpha_j$. The collection of all root vectors is called the root system, denoted as $\Delta$, and encodes all elements in the group as follows: to each $\alpha_j\in\Delta$ there corresponds a reflection $r_j(x) = x-\left(\frac{2(x|\alpha_j)}{(\alpha_j|\alpha_j)}\right)\alpha_j$, which is a reflection at the plane perpendicular to the root $\alpha_j$. The root system of $H_3$ is given by the following 30 vectors \cite{CKPS}: 
\begin{equation}\label{icosH3}
\Delta = \left\lbrace{
\begin{array}{cl}
(\pm 1,0,0) & \mbox{ and all permutations }\\
\half(\pm 1,\pm \t',\pm \t) & \mbox{ and all even permutations }
\end{array}
}\right\rbrace\,,
\end{equation}
where $\t=\half (1+\sqrt{5})$ and its algebraic conjugate $\t'=\half (1-\sqrt{5})$ are the two solutions of the quadratic equation $x^2=x+1$. 

\begin{figure}[ht]
\begin{center}
\includegraphics[width=3.8cm,keepaspectratio]{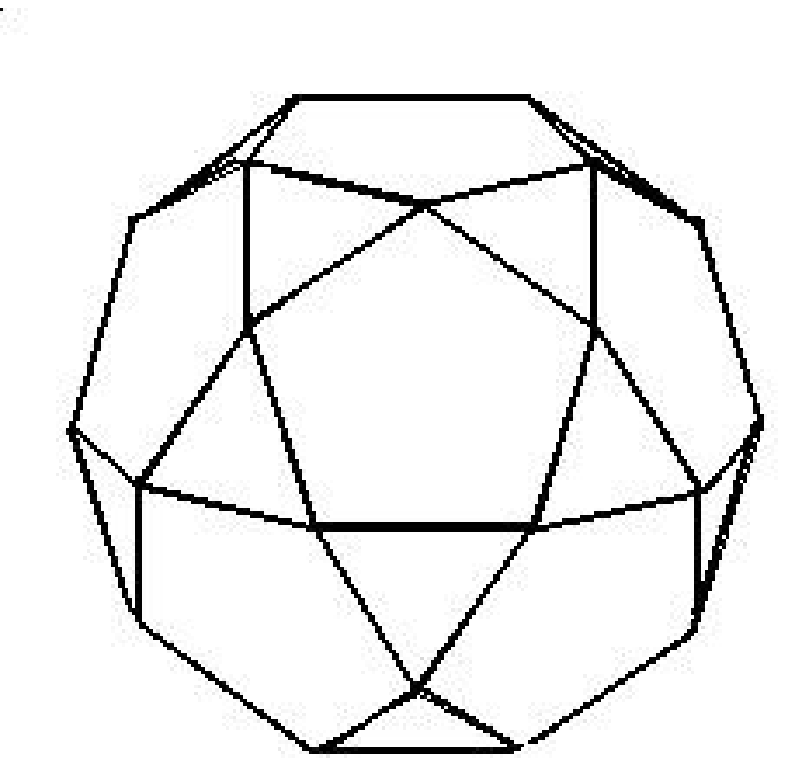}\qquad
\includegraphics[width=3.8cm,keepaspectratio]{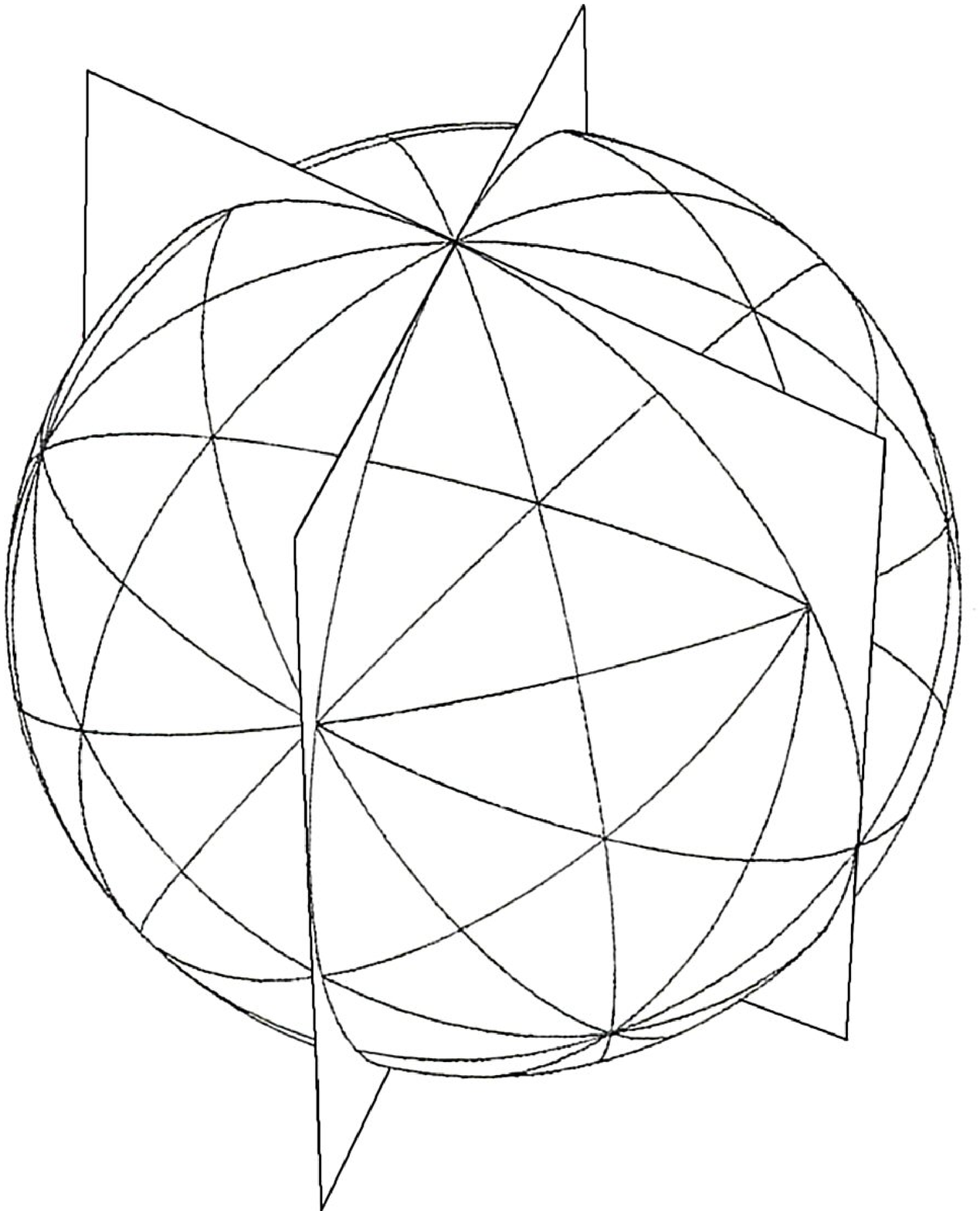}
\end{center}
\caption{Examples of planes of reflection encoded by the root vectors of $H_3$ (left) and the root polytope that encodes the locations of the planes of reflection (right).}
\label{SP}
\end{figure}

These vectors point from the centre of an icosidodecahedron to its vertices, as shown in Fig.~\ref{SP} on the left, and encode each the location of a plane of reflection, as shown on the right. 
The intersections of the planes of reflection mark the locations of the axes of rotational symmetry of the icosahedral group. For example, the intersection of the two planes in the figure corresponds to two axes of 5-fold symmetry,  which intersect the sphere at two of the 12 5-fold vertices of the (spherical) icosahedron. The other reflection planes are not shown, but their intersections with the surface of the sphere are indicated as geodesics (spherical arcs obtained as intersections of planes through the origin with the surface of the sphere). The intersections of the geodesics mark the locations of all symmetry axes, and one can hence reconstruct the locations of all 12 five-fold, 30 two-fold and 20 three-fold axes from them. 
\smallskip

In order to extend the root system in a way compatible with overall icosahedral symmetry and hence find new planes of reflection as well as the associated axes of local symmetry, one needs to observe that the information encoded by the root system $\Delta$ can further be condensed into three so-called simple roots. They form a basis from which all other root vectors can be constructed by positive or negative $\Z[\tau]$-linear combinations, where $\Z[\tau]=\lbrace a+\t b\,\vert\, a,b\in\Z\rbrace$ corresponds to an extended ring of integers. 

In particular, a possible choice of simple roots in the orthonormal basis is
\begin{equation}
\alpha_1=(0,0,1)\,,\quad\alpha_2=\tfrac12(-\t',-\t,-1)\,,\quad
\alpha_3=(0,1,0)\,.
\end{equation}

The relations between the simple roots are encoded in the Cartan matrix $C$,
\begin{equation}
C:=\left(\frac{2(\alpha_i\mid\alpha_j)}{(\alpha_j|\alpha_j)}\right)_{ij}
= 
\begin{pmatrix}
 2 &    -1 & 0 \cr 
-1 &    2  & -\t \cr
 0 & -\t &   2  
\end{pmatrix}\,.
\end{equation}

Via an affine extension this matrix is extended by an additional row and column, which encode a further root, that corresponds to an affine reflection. It has been shown in \cite{Twarock:2002AffH}, that the affine extended Cartan matrix is given by 
\begin{equation}
{\hat C} =
\begin{pmatrix}
 2    &  0  & \t' & 0     \\ 
 0    &  2  &  -1   & 0     \\
\t' &  -1 &   2   & -\t \\
 0    &  0  & -\t &   2  
\end{pmatrix}\,,
\end{equation}
and that the affine reflection act as a translation $T$ by the highest root $\alpha_H=
\t\alpha_1+2\t\alpha_2+\t^2\alpha_3= -\t'\omega_2 = (1,0,0)$. The three other reflections are cyclic operations of order two and the products of any two of them correspond to rotations around the origin: 
\begin{equation}\label{2reflect}
(r_jr_k)^M= 1 \quad\text{where}\quad
\left\{\begin{matrix}   
M=1 \quad &\text{if}\ &c_{jk}&= 2\\
M=2 \quad &\text{if}\ &c_{jk}&= 0\\
M=3 \quad &\text{if}\ &c_{jk}&=-1\\
M=5 \quad &\text{if}\ &c_{jk}&=-\t,\t'
\end{matrix}\right.
\end{equation}
The affine extended group is hence generated by the three reflections $r_1$, $r_2$, $r_3$ as well as $T$. 

Three-dimensional points sets with local symmetries can be constructed via an iterated action of the generators of the extended group on the origin. If the action of the translation operator $T$ is not restricted, space is densely filled in this way. However, if $T$ acts only a finite number of times, N say, while the action of all other operations is not restricted, point sets ${\mathcal S}(N)$ are obtained which are finite subsets of cut-and-project quasicrystals with centrally symmetric acceptance windows. For increasing $N$, the patches become larger and more dense, and correspond to cut-and-project quasicrystals with increasingly larger acceptance windows. Examples have been worked out explicitly for the two-dimensional subgroup $H_2$ in \cite{Twarock:2002AffH}, and for $H_3$ in \cite{Twarock:2002Onion}, where applications to carbon cage structures have been considered.  

Due to the fact that that $T$ acts as a translation by the highest root, a simple expression for these point sets can be given in terms of the root system $\Delta$ (see \cite{Twarock:2002Onion} for the case of $H_3$): 
\begin{equation}\label{SN}
{\mathcal S}(N) := \left\lbrace  \sum_{\alpha \in\Delta} n_\alpha \alpha \,\vert \,  n_\alpha \in \N\cup\lbrace 0 \rbrace, \sum_{\alpha \in\Delta} n_\alpha \leq N   \right\rbrace\,,
\end{equation}
i.e. ${\mathcal S}(N)$ consists of all points that are obtained as an $\N\cup\lbrace 0 \rbrace$-linear combination of up to $N$ roots in $\Delta$. These sets form the starting point for the construction of icosahedral particles with local symmetries in the next subsection. 

\subsection{The tool-kit}\label{twotwo}

The sets ${\mathcal S}(N)$ in (\ref{SN}) decompose into disjoint subsets ${\mathcal S}_k(N)$ of vectors of equal length that are invariant under the action of $H_3$:   
\begin{equation}\label{orbit}
{\mathcal S}(N) = \cup_{k=1}^{K_N}{\mathcal S}_k(N) \,,
\end{equation}
where $K_N$ denotes the number of different subsets of vectors of equal length in ${\mathcal S}(N)$. These subsets are of different cardinalities, and a given set ${\mathcal S}_k(N)$ may increase in cardinality for increasing $N$, i.e.~one has the set inclusion ${\mathcal S}_k(N) \subseteq {\mathcal S}_k(N+l)$ for $l\in\N$. 

Each set ${\mathcal S}_k(N)$ encodes the surface structure of a particle that has local symmetry axes compatible with the overall group structure. Since only the orientations but not the length of the vectors in the set is important for the reconstruction of the locations of the local reflection planes, two sets ${\mathcal S}_k(N)$ correspond to the same particle if they contain the same vectors modulo length. Therefore the range of different (modulo length) sets ${\mathcal S}_k(N)$ corresponds to the spectrum of different icosahedral particles with local symmetries. The tool-kit given here provides a method for the systematic evaluation of this information and hence for the construction of the surface structures of all such particles. 
\medskip

By definition, ${\mathcal S}(1)$ corresponds to the root system $\Delta$. The reflection planes orthogonal to the roots intersect precisely at the symmetry axes corresponding to the (global) symmetries of the icosahedral group, and it is hence a structure without local symmetries. Therefore, local symmetries are encoded in the point sets ${\mathcal S}(N)$ with $N\geq 2$. For each $N$, the sets ${\mathcal S}_k(N)$ in (\ref{orbit}) need to be evaluated individually. This is done by an inductive process starting with the case $N=2$. It is presented in the form of a tool-kit which facilitates easy computational implementation and can be used also by people not familiar with the group theoretical set-up: 

\begin{enumerate}
\item Add any two vectors in (\ref{icosH3}) and group these vectors into sets ${\mathcal S}_k(2)$, $k=1,\ldots, K_2$, of vectors of equal length. For each $k$, compute the planes through the origin that are perpendicular to the vectors in ${\mathcal S}_k(2)$ according to the equation 
\begin{equation}\label{symplane}
\sum_{j=1}^{3} v_j x_j =0
\end{equation}
for any vector $v\in {\mathcal S}_k(2)$ with Cartesian components $v_j$. The intersections of these planes determine the locations and types of the corresponding local symmetry axes. They may either be computed analytically or graphically displayed with a visualisation tool such as Maple. 
\item Compute the structure of ${\mathcal S}(N)$ inductively based on the structure of ${\mathcal S}_k(N-1)$ as follows: Add any N vectors in (\ref{icosH3}) and group them by their length into sets ${\mathcal S}^j(N)$. If the length of the vectors in a set ${\mathcal S}^j(N)$ corresponds to that of one of a sets at iteration step $N-1$, ${\mathcal S}_k(N-1)$ say, form a new set that contains both sets, i.e. introduce a new set ${\mathcal S}_k(N):={\mathcal S}^j(N)\cup {\mathcal S}_k(N-1)$.  Otherwise, add the set as a new set to the list, i.e. ${\mathcal S}_l(N)={\mathcal S}^j(N)$, where the index $l$ is chosen such that sets are numbered consecutively. The collection of sets obtained in this way corresponds to the sets ${\mathcal S}_k(N)$, $k=1,\ldots, K_N$ in (\ref{orbit})\footnote{As mentioned before, two sets of vectors that are identical modulo length represent identical particles. However, none of them can be discarded from the list as they may lead, at a later iteration step, to different point sets when further points are concatenated to them.}. For each of them, compute the corresponding symmetry planes according to (\ref{symplane}) as before. 
\end{enumerate}

\begin{figure}[ht]
\begin{center}
\includegraphics[width=5cm,keepaspectratio]{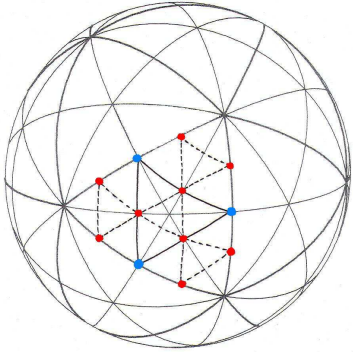}
\end{center}
\caption{Three axes of local symmetry shown in blue, additional symmetry axes added to the set for larger $N$ are shown in red.}
\label{TRIANGLES}
\end{figure}

Applications of this tool-kit are discussed in section \ref{three}. 
Note that the intersections of the reflection planes with the surface of the sphere lead to spherical tessellations of the surface of the sphere. The set inclusions of the type ${\mathcal S}_k(N_1)\subset {\mathcal S}_k(N_2)$ for $N_2>N_1$ imply that the spherical tiling (tessellations) corresponding to these sets are related by {\it selfsimilarity}: larger copies of the tiles (areas) implied by the  spherical tiling related to ${\mathcal S}_k(N_1)$ are replaced by an integer number of rescaled versions thereof in the spherical tiling related to ${\mathcal S}_k(N_2)$. An example is shown in Fig.~\ref{TRIANGLES}, where each of the four triangles in the original triangulation (with vertices shown as blue dots) has been replaced by four rescaled (smaller) versions in the refined triangulation. 

The tiling is shown only in one of the 20 faces of the spherical icosahedron (or, equivalently, in three copies of the fundamental domain of the icosahedral group), because the remainder of the tiling follows from this via the actions of the icosahedral group. Planes represented by lines connecting the blue dots arise from the vectors in the set ${\mathcal S}_k(N_1)$, while planes represented by dashed lines connecting the red dots correspond to the vectors that are contained in 
${\mathcal S}_k(N_2)$ in addition. 

This phenomenon is also known from the study of quasicrystals \cite{Senechal:1996}. Since quasicrystals are obtained from a projection formalism that is related to this construction as discussed in the introduction, this is not surprising. 

\section{Applications to viruses}\label{three}

In this section it is demonstrated how the theory can be applied in virology. In subsection \ref{threeone} the tool-kit is applied to derive the surface structures of new viral particles with local symmetries. In subsection \ref{threetwo} it is shown that the theory answers, as a by-product, an open question in virology: it explains why only certain types of capsomeres (protein clusters) are observed in viral capsids. 

\subsection{New particles with local symmetries}\label{threeone}

The local symmetries of particles with icosahedral symmetry can be obtained from the root system of the non-crystallographic Coxeter group $H_3$ via the tool-kit presented in the previous section.  
According to this, we start by considering the set ${\mathcal S}(2)$ of all vectors obtained via a linear combinations of any  two root vectors in (\ref{icosH3}) and decompose this set into subsets ${\mathcal S}_k(2)$ of roots of equal length. The types of the local symmetry axes that are obtained as intersections of the reflection planes encoded by the vectors in a set ${\mathcal S}_k(2)$ depend on the relative orientations of these vectors, as can be seen from the examples discussed below. 

\begin{enumerate}
\item {\it A particle with local 3-fold axes:}

Consider the subset ${\mathcal S}_k(2)$ of ${\mathcal S}(2)$ given by linear combinations of any two vectors pointing to vertices of the icosidodecahedron with a joint edge, such as, for example, the vertices shown in blue in Fig.~\ref{CK}. The resulting vectors intersect the edges at the points shown in red. In order to reconstruct the locations of the reflection planes orthogonal to these vectors, note that the organisation of the edges reflects the subgroup structure of $H_3$: they are grouped in decagonal arrangements according to copies of the subgroup $H_2$. Each such decagon encodes 10 root vectors of $H_3$, which in turn encode the symmetry planes that intersect at a global 5-fold vertex and its antipode on the sphere.  Therefore, the bisection of the edges in the root vector picture in Fig.~\ref{CK} on the left correspond to symmetry planes that bisect the angles between two planes meeting at a 5-fold vertex at an angle $2\pi / 10$. The geodesics corresponding to the intersection of such planes with the sphere are shown in Fig.~\ref{CK} on the right superimposed on the spherical icosahedron. In order to keep the presentation transparent, only the geodesics in  one of the 20 faces of the spherical icosahedron are shown: geodesics arising from bisections of the angles within this spherical triangle are shown as dashed lines; moreover, the geodesics arising from bisections of angles in different spherical triangles (faces) are shown as dotted lines in order to show their impact on the spherical triangle under consideration. 
\begin{figure}[ht]
\begin{center}
\includegraphics[width=5cm,keepaspectratio]{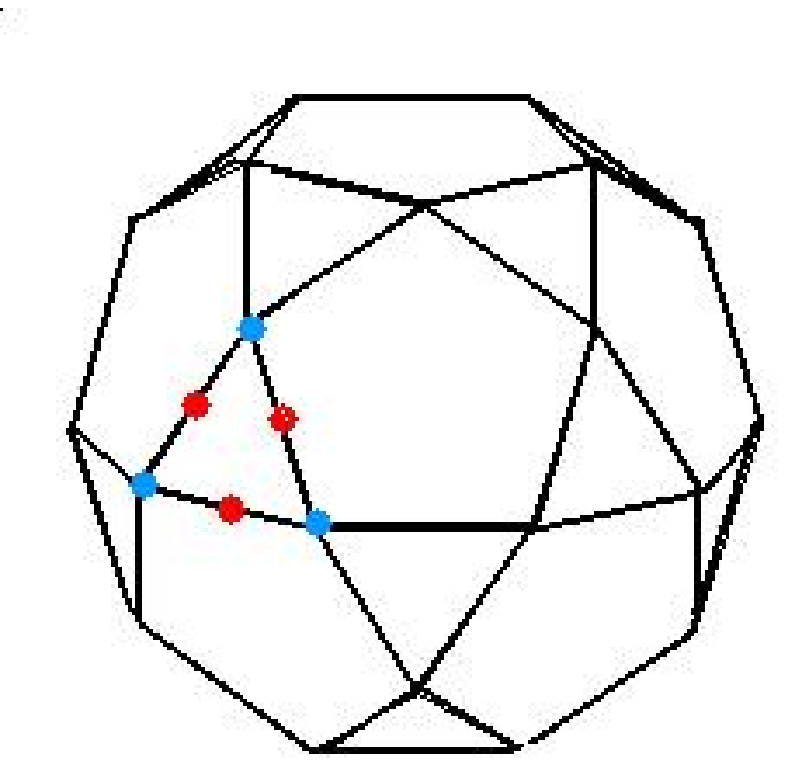}\qquad
\includegraphics[width=5cm,keepaspectratio]{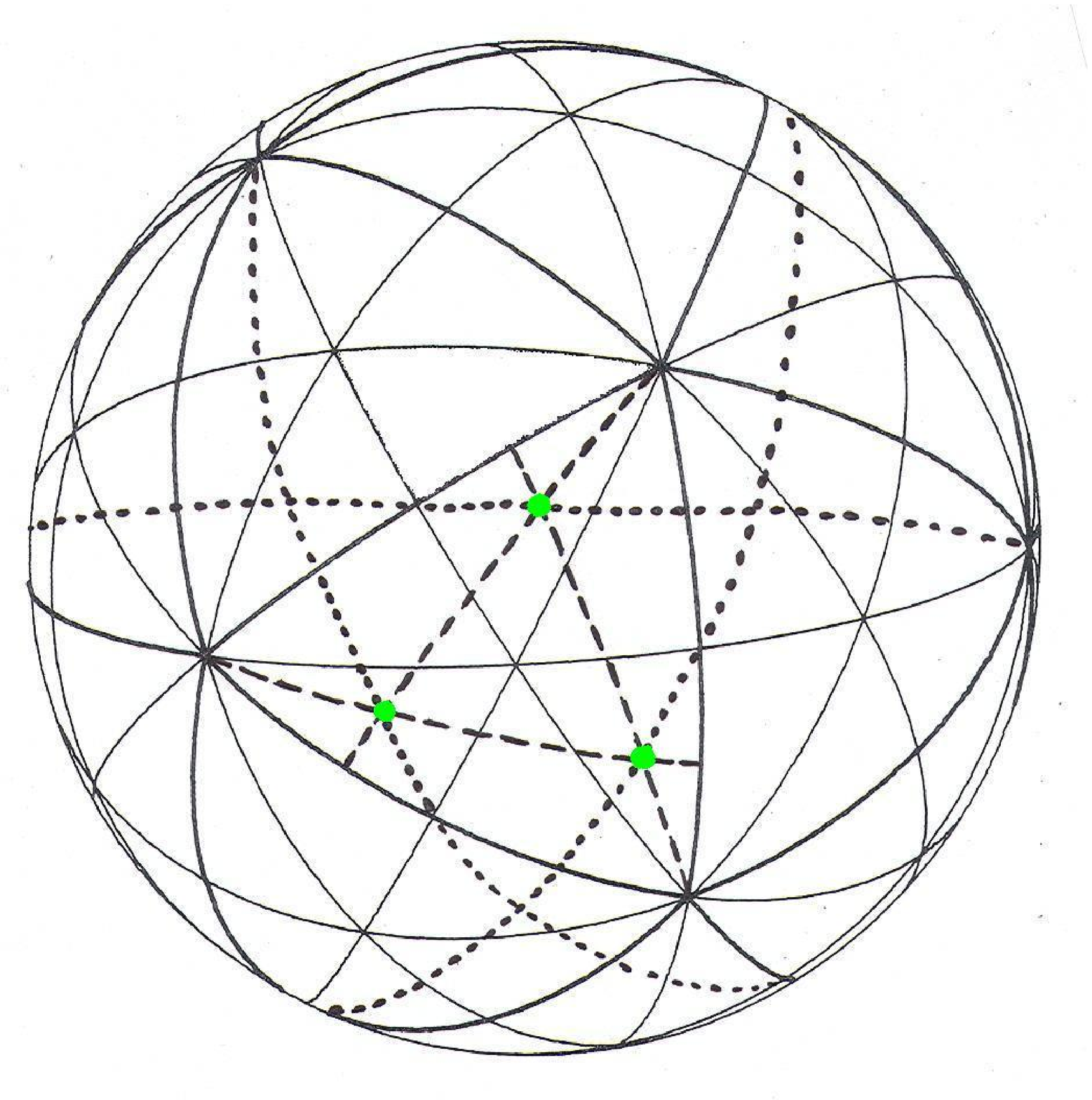}
\end{center}
\caption{Linear combinations in the root system (left) leading to a spherical tiling with local 6-fold symmetry axes at the green vertices (right).}
\label{CK} 
\end{figure}

Note that the intersections of these geodesics determine the locations of the local symmetry axes: there are three local symmetry axes of order six in the spherical triangle. Since this spherical triangle contains three copies of the fundamental domain of the icosahedral group, this setting corresponds to one new local 6-fold symmetry axis that is reproduced 60 times under the action of the icosahedral group. Note that the planes of symmetry, respectively the geodesics corresponding to them, not only fix the locations of the symmetry axes but also the relative orientation of the planes of symmetry and hence the surface structure of the corresponding viral capsids is completely determined by this formalism. 

\item{\it A particle with local 5-fold axes:}

In order to obtain a particle with local 5-fold symmetry axes, a different subset ${\mathcal S}_k(2)$ of ${\mathcal S}(2)$ has to be considered, which inherits its order of symmetry from the global 5-fold axes. For this, consider linear combinations of root vectors that are located on two different $H_2$-orbits, as for example the vectors pointing to the vertices shown in blue in Fig.~\ref{RT1} on the left. These vertices are located on two $H_2$-orbits that are shown in green and red, respectively, and intersect at the vertex shown in yellow. Fig.~\ref{RT1} on the right shows how this picture translates into reflection planes. In particular, part of the geodesics corresponding to the reflection planes perpendicular to the 10 root vectors pointing to vertices on the green, respectively red, $H_2$-orbit are shown schematically in green, respectively red, around 5-fold vertices in the spherical tiling. The geodesics corresponding to the two root vectors pointing to vertices marked by blue dots in the figure on the left are shown in blue in the spherical tiling on the right. Fig.~\ref{RT1} hence provides a dictionary on how to translate the geometry of the root vectors into geodesics on the sphere. 
\begin{figure}[ht]
\begin{center}
\includegraphics[width=5cm,keepaspectratio]{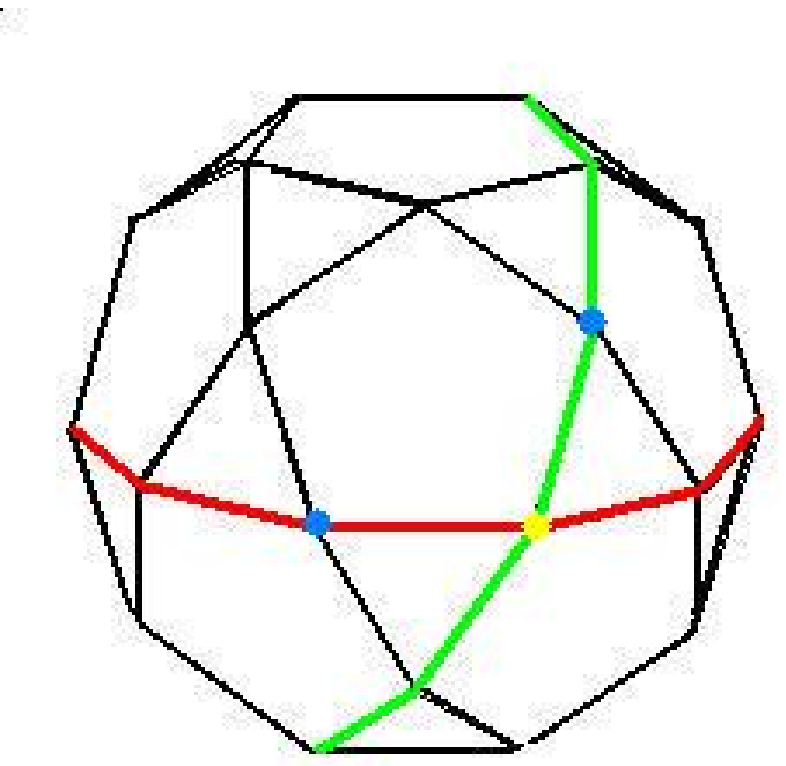}\qquad
\includegraphics[width=5cm,keepaspectratio]{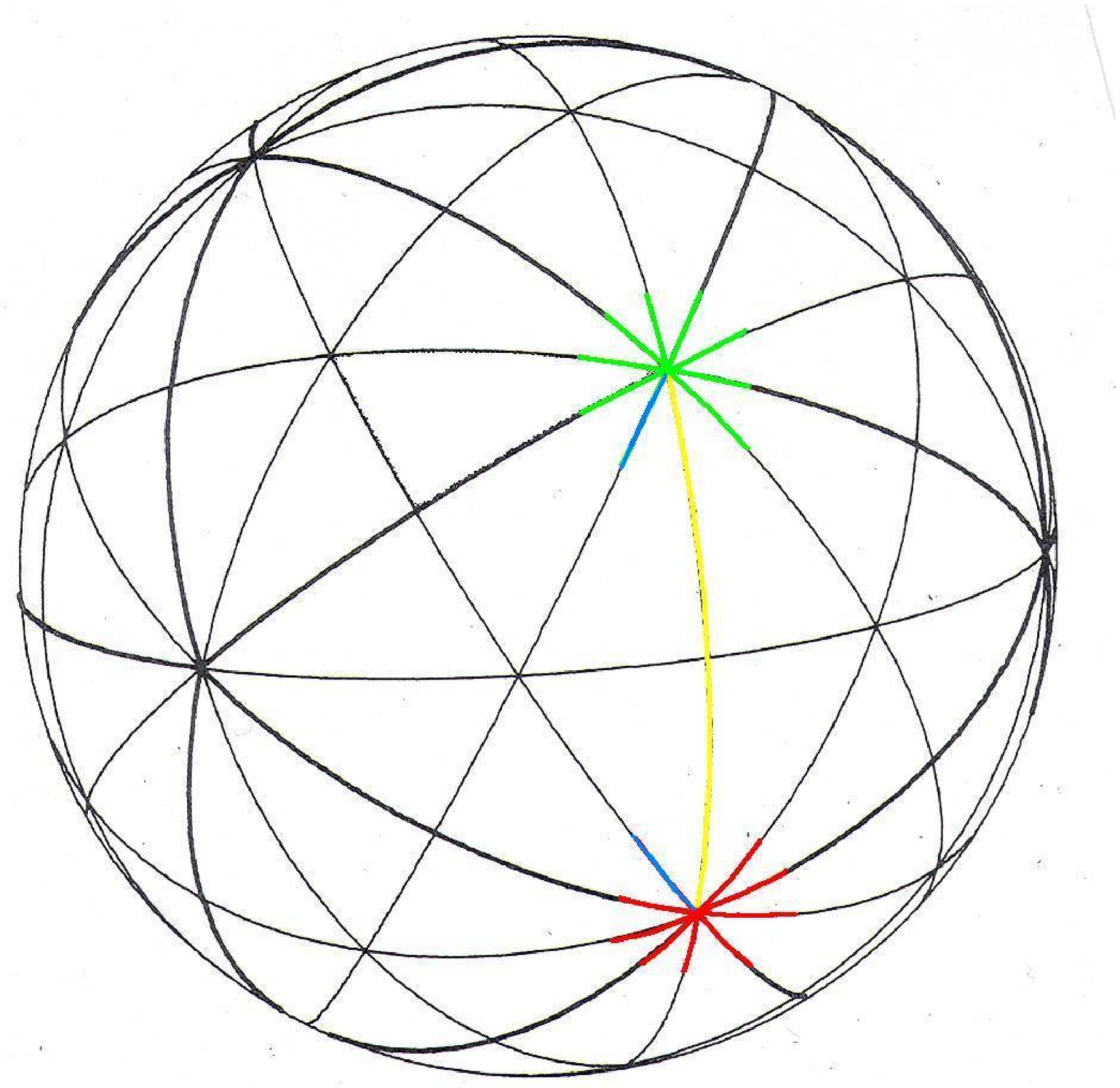}
\end{center}
\caption{A dictionary for translating the $H_2$-orbit structure in the root system (left) into geodesics corresponding to reflection planes (right).}
\label{RT1}
\end{figure}

Based on this, it is possible to construct the axes of local five-fold symmetry as follows. The linear combinations of any two root vectors corresponding to the blue dots in Fig.~\ref{RT2} (left) corresponds to a vector that points from the centre of the icosidodecahedron to any of the vertices marked in red in Fig.~\ref{RT2} (left). Reconstructing the corresponding planes of symmetry as before leads to a network of geodesics with intersections of order 10 rather than 5.  However, in order to obtain the spherical tiling that represents the surface structure of the viral particle, note that also the global 5-fold axes appear as 10-fold axes in this representation, because the angles between the five reflection planes at each 5-fold vertex are bisected by the reflection planes originating at the 5-fold vertex at its antipode on the sphere. Similarly, the local symmetry axes have to be interpreted as 5-fold symmetry axes in disguise, and every second geodesic has to be ignored. This leads to two different spherical tilings, shown in Fig.~\ref{RT2} in the middle and on the right, respectively.  
\begin{figure}[ht]
\begin{center}
\includegraphics[width=4.8cm,keepaspectratio]{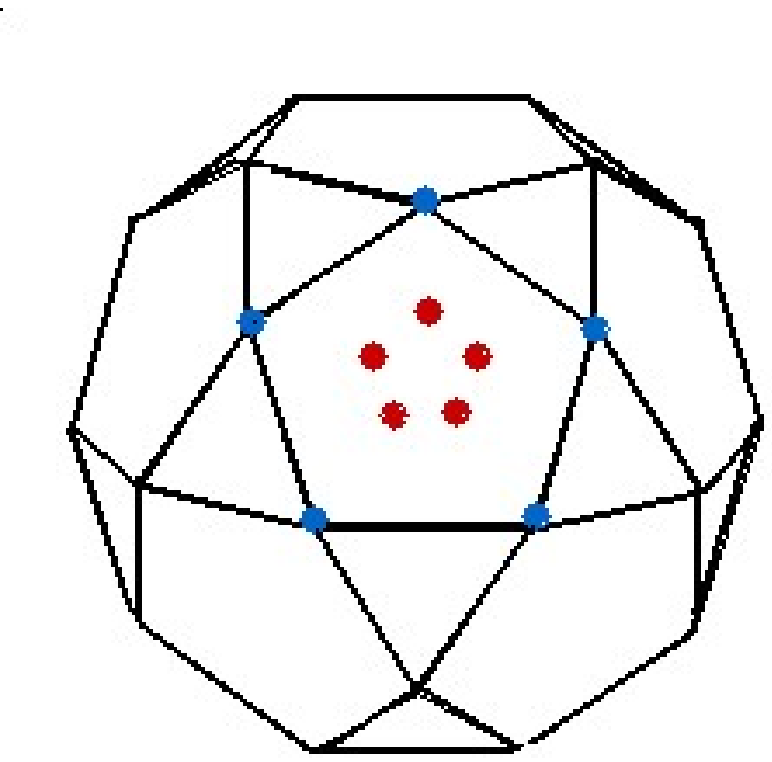}
\includegraphics[width=4.8cm,keepaspectratio]{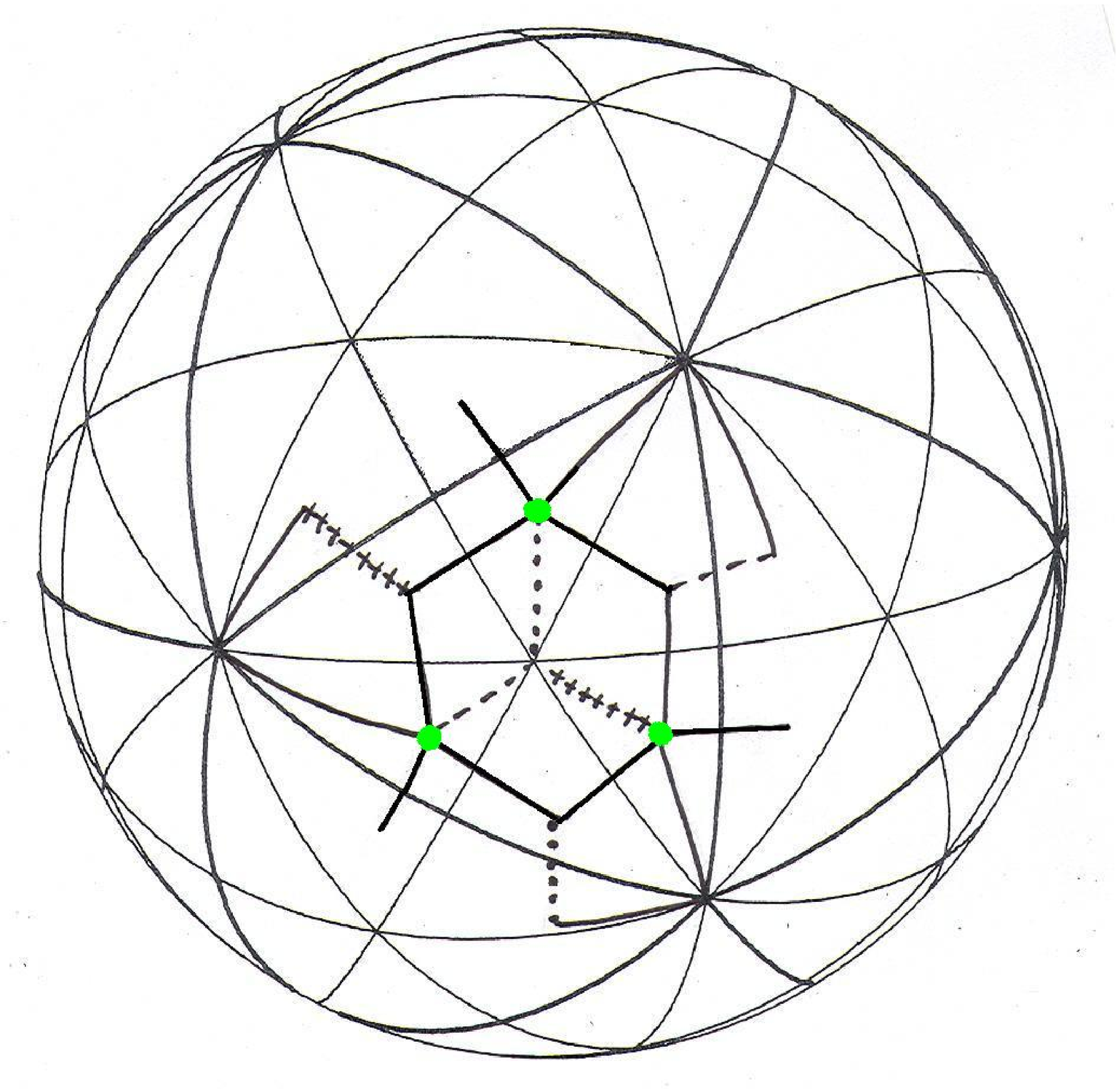}
\includegraphics[width=4.8cm,keepaspectratio]{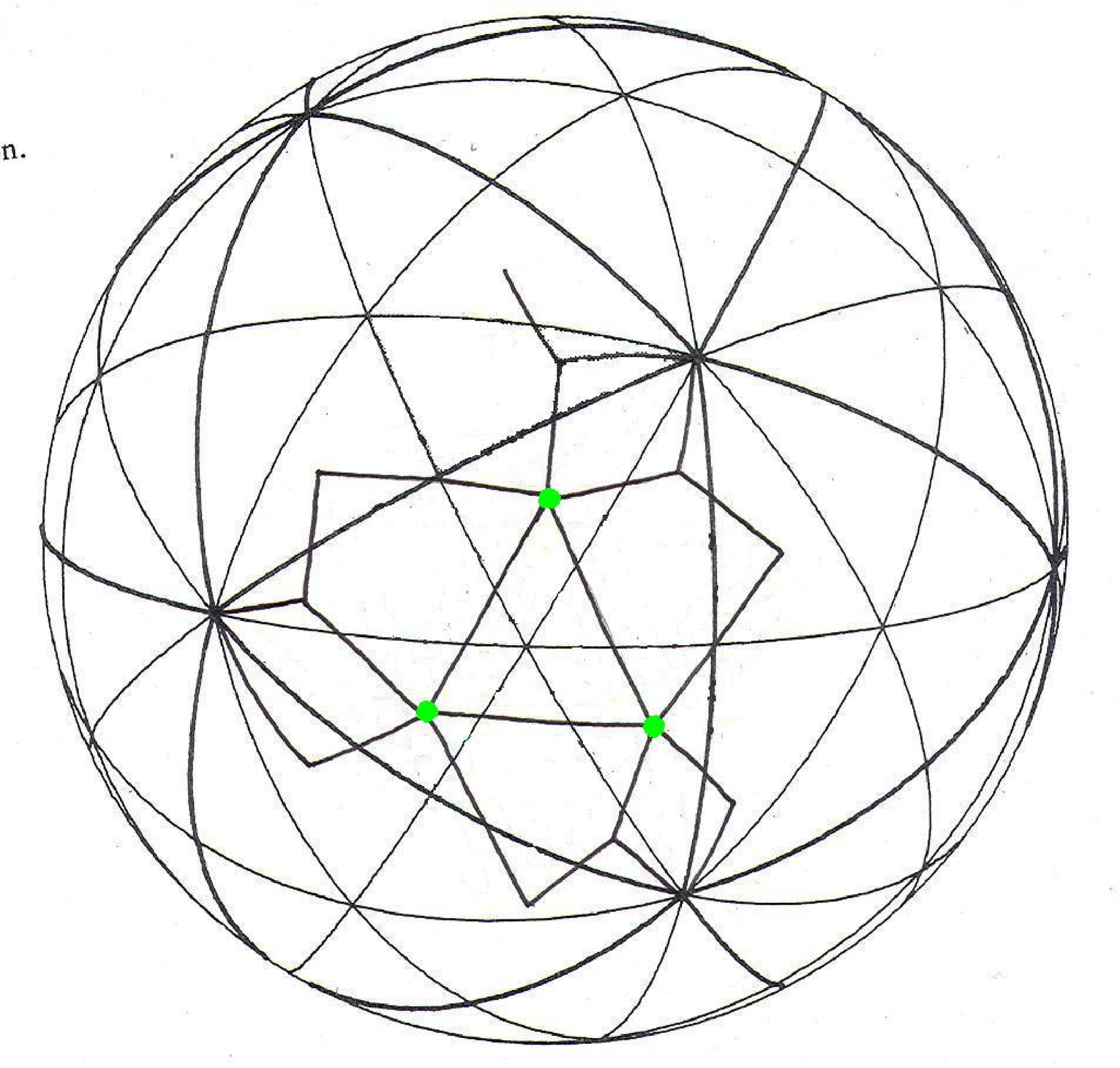}
\end{center}
\caption{Linear combinations in the root system (left) leading to two spherical tilings with local 5-fold symmetry axes at the green vertices (middle and right).}
\label{RT2}
\end{figure}

In order to visualise the structure of the viral capsids encoded by these tilings, the rules of the tiling approach need to be used \cite{Twarock:2005a}: protein subunits are located in the corners of the tiles that meet at the local symmetry axes and are represented schematically as dots. The colours of the dots encode which proteins are mapped onto each other under the action of the overall symmetry group; the proteins themselves are identical and hence indistinguishable. Tiles with two and three dots represent, respectively, dimer- and trimer-interactions, i.e. interactions between, respectively, two and three protein subunits. They are visualised schematically as spiral arms and may correspond for example to C-terminal arm extensions of the proteins. With this interpretation, the tiling in Fig.~\ref{RT2} on the right leads to the surface structure shown in Fig.~\ref{surfaces}. 
\begin{figure}[ht]
\begin{center}
\includegraphics[width=4.8cm,keepaspectratio]{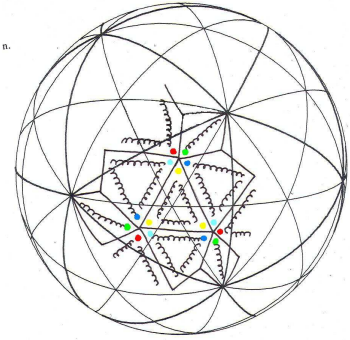}
\end{center}
\caption{Schematic representation of the surface structures encoded by the tiling in Fig.~\ref{RT2} (right).}
\label{surfaces}
\end{figure}

Note that the predictions of this tiling differ both by the orientations of the protein clusters (pentamers), which are rotated by an angle of $2\pi/5$ with respect to the other alternative, and by the types and locations of the intersubunit bonds that stabilize the capsid. The tiling in the middle of Fig.~\ref{RT2} corresponds to the tiling in \cite{Twarock:2004a} and represents viral particles in the family of Papovaviridae. Viral particles following the alternative in Fig.~\ref{surfaces} are not known at present. 

\end{enumerate}

Note that these sets are only two examples of sets in the decomposition of ${\mathcal S}(2)$. They have been chosen because they are the most instructive examples. There exists, for example, also a subset that does not introduce any local symmetry axes: it corresponds to linear combinations of each root vector with itself. These vectors point to the vertices of a larger copy of the icosidodecahedron, and the set hence encodes only the global symmetry axes (like the root system itself) and hence contains  no new information. 

While the geometry of the particle in item 1 is known already from the Caspar-Klug classification (it belongs to a $T=7d$ capsid), the geometries in item 2 are new. They correspond to particles with only 5-fold axes of symmetry (global and local). Due to overall icosahedral symmetry, any local 5-fold axis occurs in multiples of 60, so that they are indeed the smallest geometries corresponding to particles with local five-fold axes. A classification of all icosahedral particles with local symmetries  based on the theory developed here is in progress  \cite{Twarock:2005c}.

\subsection{Orders of the capsomeres}\label{threetwo}

The centres of the capsomeres in viral capsids correspond to the local and global axes of symmetry. Therefore, the spectrum of different capsomeres that may occur in nature is determined completely by the orders of the symmetry axes that can occur from a mathematical point of view. The group theoretical formalism developed here implies that the orders of all local symmetry axes must correspond to multiples of those of the global ones, because they inherit their symmetry from the structure of the root system due to the fact that they arise from $\Z$-linear combinations of the root vectors. 
Therefore, the orders of the local symmetry axes must be $\Z$-multiples of  2, 3 or 5, and local symmetry axes of, for example, order 7 cannot occur. Moreover, since smaller particles are obtained from linear combinations of only a few root vectors, multiples of more than twice the order of the global axes do not occur for them. Hence, for all smaller  icosahedrally symmetric particles, which form the majority of viruses observed to date, one would expect a spectrum of local symmetries of order 3, 5, 6 and 10. All of these are known to occur: order three has been observed for bacteriophage MS2 \cite{Valegard:1990}, order five for polyomavirus \cite{Rayment:1982} and Simian Virus 40 \cite{Liddington:1991}, order six in all Caspar-Klug cases, and order 10 for L-A virus of yeast \cite{Caston:1997}. All these cases follow the construction principle outlined above. 

\section{Discussion}

The formalism presented here is used in \cite{Twarock:2005c} to describe the surface structures of viral capsids in terms of tilings. The vertex sets of the tilings contain the points, that correspond to the local and local symmetry axes, and mark the locations of the capsomeres. The edges of the tiles follow the geodesics that arise as intersections of the planes of (local and global) symmetry with the surface of the sphere. In this sense, the tiling picture and the group theoretical picture in terms of local symmetry axes are equivalent ways of modelling the surface structures of viral capsids. 

This may suggest that the tiles are of secondary importance and that the symmetry axes should be the focus of attention. However, in this way, important additional information on the structure of viral capsids would be neglected    as tiles represent the interactions between subunits belonging to different capsomeres and are hence biologically meaningful objects. In particular, the measures and shapes of the tiles  are correlated with the strengths and types of  the intersubunit bonds they represent \cite{Twarock:2005c}. Their structure is hence very important information in order to determine the types of the interactions (i.e. whether they are trimer- or dimer-interactions, that is interactions between three, respectively two, protein subunits), as well as their exact locations. This information is crucial input for the construction of assembly models \cite{KTT,KMT}, that investigate how viral capsids are assembled from their building  blocks, as well as for the modelling and classification of additional structures  such as the covalent bonds responsible for crosslinking \cite{Twarock:2005b}. 

\section*{Acknowledgements}
Financial support via an EPSRC Advanced Research Fellowship is gratefully acknowledged.

\end{document}